\documentclass[aps,prl,amsmath,amssymb,floatfix,twocolumn,  superscriptaddress]{revtex4}
\usepackage{multirow}
\usepackage{bbold}
\usepackage{graphicx}
\usepackage{subfigure}
\usepackage{color}
\usepackage{hyperref}

\usepackage{amsfonts, relsize, color}
\usepackage{graphics}
\usepackage{graphicx}
\usepackage{subfigure}
\usepackage{hyperref}
\usepackage{color}

\begin{document}
\title{Universal optical conductivity of a disordered Weyl semimetal}

\author{Bitan Roy}
\affiliation{Condensed Matter Theory Center and Joint Quantum Institute, University of Maryland, College Park, Maryland 20742-4111, USA}

\author{Vladimir Juri\v ci\' c}
\affiliation{Nordita,  Center for Quantum Materials,  KTH Royal Institute of Technology and Stockholm University, Roslagstullsbacken 23,  10691 Stockholm,  Sweden}

\author{Sankar Das Sarma}
\affiliation{Condensed Matter Theory Center and Joint Quantum Institute, University of Maryland, College Park, Maryland 20742-4111, USA}

\begin{abstract}
Topological Weyl semimetals, besides manifesting chiral anomaly, can also accommodate a disorder-driven unconventional quantum phase transition into a metallic phase. A fundamentally and practically important question in this regard concerns an experimentally measurable quantity that can clearly distinguish these two phases. We show that the optical conductivity while serving this purpose can also play the role of a bonafide order parameter across such disorder-driven semimetal-metal quantum phase transition by virtue of displaying distinct scaling behavior in semimetallic and metallic phases, as well as inside the quantum critical fan supporting a non-Fermi liquid. We demonstrate that the correction to dielectric constant and optical conductivity in a dirty Weyl semimetal due to weak disorder is independent of the actual nature of point-like impurity scatterers. Therefore, optical conductivity can be used as an experimentally measurable quantity to study the critical properties and to pin the universality class of the disorder-driven quantum phase transition in Weyl semimetals.
\end{abstract}

\flushbottom
\maketitle
\thispagestyle{empty}

\emph{Introduction}: Understanding and characterizing phase transitions is one of the most important problems in condensed matter physics.
Identification of distinct phases of matter and the possible phase transitions among them necessarily rely on the existence of a physical quantity that behaves differently in two phases and as such can potentially serve as a bonafide \emph{order parameter} (OP) across the transition. The notion of an OP is as germane near a zero-temperature quantum phase transition (QPT), driven by quantum fluctuations, as near the finite-temperature classical phase transition, where thermal fluctuations dominate~\cite{sachdev, herbut}. With increasing complexity of various phases, the horizon of OPs has expanded enormously, and topological OPs, which globally characterize a phase of matter, have recently emerged~\cite{TI-review-1, TI-review-2}. Moreover, the landscape of topological states has been extended to gapless systems featuring quasiparticles at arbitrarily low energies in the bulk, with Weyl semimetal (WSM), discovered in various three-dimensional gapless semiconductors~\cite{taas-1, tasas-2, taas-3,nbas-1,tap-1,nbp-1, nbp-2,tas,borisenko, chiorescu},  standing as the paradigmatic representative. The constituting Weyl nodes are topologically protected and act as a source (monopole) and a sink (anti-monopole) of Berry flux in the momentum space, manifesting through Adler-Bell-Jackiw chiral anomaly and surface Fermi arcs~\cite{burkov-review,das-rmp}.

\begin{figure}[htb]
\includegraphics[width=8.cm, height=6.5cm]{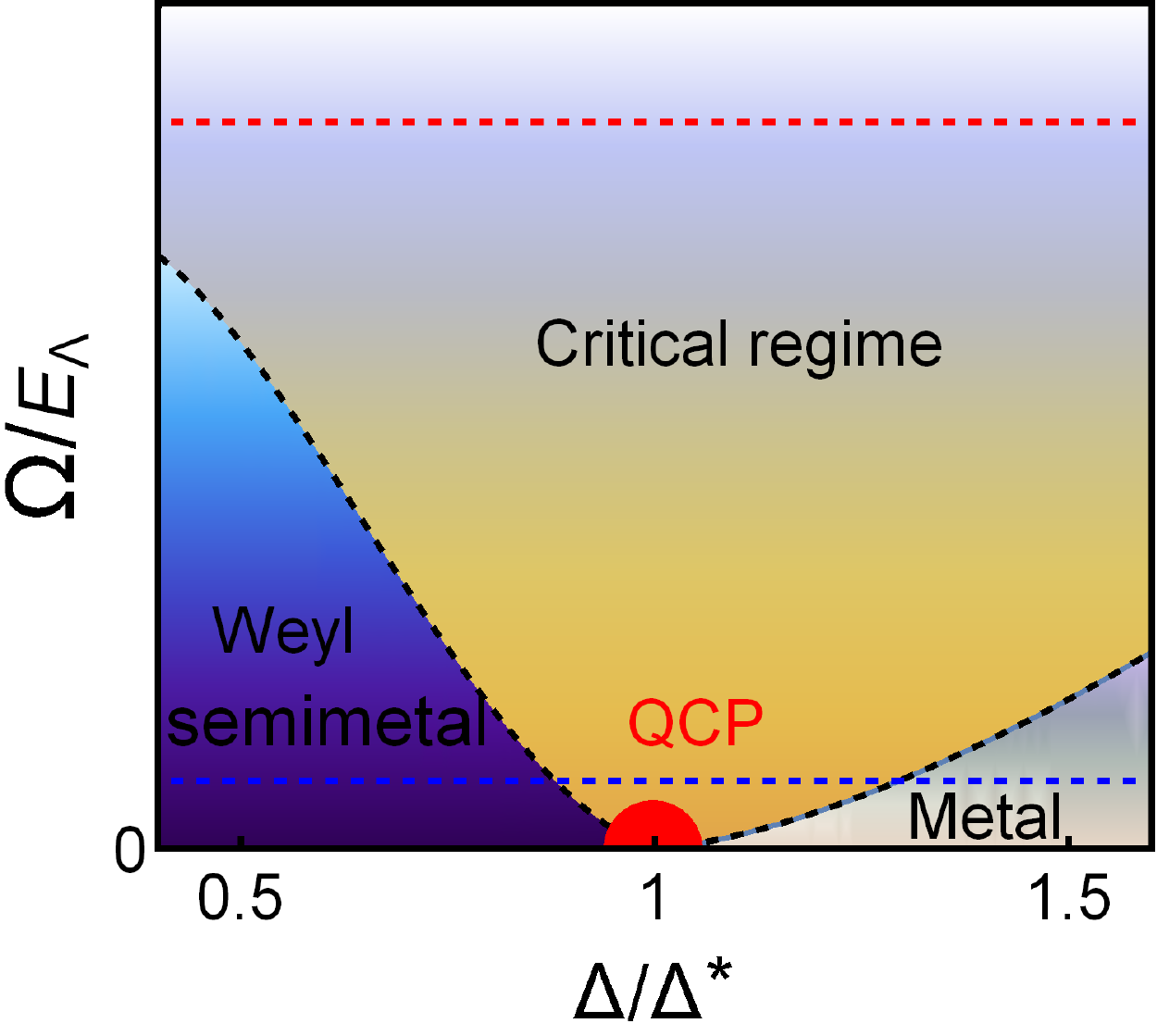}
\caption{(Color online) A schematic phase diagram of a dirty Weyl semimetal at finite frequencies ($\Omega$), subject to random charge impurities, where $E_\Lambda \sim v \Lambda$ is the ultraviolet cutoff for energy. All the phases and the quantum critical point (red dot)  exist only at zero frequency. Various crossover boundaries (black lines), such as the ones between the critical regime and Weyl semimetal or metal, have been estimated from the scaling of specific heat at finite temperatures~\cite{pixley-2} and average density of states at finite energies~\cite{roy-bera}. The red line marks the high energy cut-off above which the continuum description of a WSM based on linearly dispersing quasiparticles breaks down. Blue line shows the location of Fermi energy (often unknown). WSM-metal QPT is tuned by disorder ($\Delta$) and takes place at a critical strength of disorder $\Delta=\Delta^\ast$ (see text).  The optical conductivity inside the Weyl semimetal and critical regime respectively scales/vanishes as $\Omega$ and $\Omega^{1/z}$, while it becomes finite in the metallic phase as $\Omega \to 0$. As frequency is increased optical conductivity displays smooth crossovers between distinct regimes (represented by color gradient in the phase diagram). In the vicinity of the WSM-metal QCP at $\Delta=\Delta^{\ast}$, the phase boundary between the critical regime and WSM or metal scales as $\delta^{\nu z}$ (see text). In the plot we use  the one-loop result for the critical exponents $\nu=1$ and $z=3/2$ at the QCP corresponding to the QPT driven by the potential or axial chemical potential disorder. }\label{criticalregime}
\end{figure}

In addition to its topological properties, WSM can also support a disorder-tuned unconventional QPT toward a diffusive metallic phase at a finite disorder strength~\cite{fradkin, shindou, ominato, chakravarty, roy-dassarma, radzihovsky, altland,imura, herbut-disorder, brouwer-1, pixley-1, brouwer-2, pixley-2, ohtsuki, roy-bera, hughes}, see Fig.~\ref{criticalregime}, and we propose that optical conductivity (OC) can expose the rich phase diagram of a dirty WSM at finite frequencies. Unveiling such novel quantum critical phenomena in real materials, however, necessarily encounters technical difficulties. For example, the average density of states at the Weyl points, although has been proposed as a possible OP across such semimetal-metal QPT~\cite{herbut-disorder, pixley-1, pixley-2, ohtsuki, roy-bera}, its measurement through compressibility in three-dimensional systems is extremely challenging, and may become even more complicated due to unwanted but likely presence of charged puddles~\cite{puddle-1, dassarma-puddle}, Lifshitz tail and rare region effects~\cite{nandkishore, pixley-rareregion} in vicinity of the Weyl nodes, as well as due to pinning of the Fermi energy away from the Weyl points. These mechanisms can musk the WSM-metal quantum critical point (QCP)~\cite{puddle-1, dassarma-puddle} or perhaps even convert it into a hidden QCP~\cite{nandkishore, pixley-rareregion}, therefore demanding the search of a measurable quantity that can unearth the underlying QCP by exposing the wide quantum-critical regime away from the pristine QCP [see Fig.~\ref{criticalregime}]. While its inter-band component is capable of bypassing these barriers, we show here that the OC can also display a single parameter scaling across the WSM-metal QPT, thus being suitable as a promising candidate for an experimentally viable OP in a dirty WSM.

We establish that while the OC vanishes linearly with frequency ($\Omega$) in a clean WSM, weak disorder leads to a nontrivial but \emph{universal} (up to a sign) correction, irrespective of the actual nature of elastic scatterers. Thus both clean and weakly disordered WSMs behave as \emph{power-law insulators}. It is worth mentioning that OC ($\sigma$) has been experimentally measured in the three-dimensional materials featuring linearly dispersing quasiparticles in the bulk, suggesting that $\sigma(\Omega) \sim \Omega$~\cite{ueda, basov, drew}. On the other hand, in the metallic phase the zero-frequency OC becomes finite, displaying a universal power-law dependence on disorder strength (measured from the critical one), set only by the correlation length exponent ($\nu$). Inside the quantum critical regime, constituted by disorder-induced strongly coupled gapless critical modes supporting a non-Fermi liquid, the OC vanishes as $\sigma(\Omega) \sim \Omega^{1/z}$ when frequency $\Omega \to 0$, with $z$ being the dynamic scaling exponent that together with correlation length exponent ($\nu$) defines the universality class of the WSM-metal QPT. As we show here, measurement of OC in a wide frequency range in fact offers a unique opportunity to unearth the universality class of the WSM-metal QPT, besides exposing a rich phase diagram of dirty WSMs. Although we here analyze the scaling behavior of OC at $T=0$, its jurisdiction covers the entire \emph{collisionless} regime ($\Omega \gg T$).

\emph{Model}: Quintessential properties of a WSM can be captured by the effective low-energy Hamiltonian
\begin{equation}\label{weylcontinuum}P
H_0 = \hbar v \; \int \frac{d^3 \mathbf{p}}{(2 \pi)^3} \; \Psi^\dagger_{\mathbf{p}} \left( \Gamma_1 p_1 + \Gamma_2 p_2 + \Gamma_3 p_3 \right)\Psi_{\mathbf{p}},
\end{equation}
where $v$ is the the Fermi velocity of Weyl fermions, assumed to be isotropic for simplicity, and $p_j$ are components of momentum. Three mutually anticommuting matrices are defined as $\Gamma_j=\tau_3 \otimes \alpha_j$ with two sets of Pauli matrices $\boldsymbol \tau$ and $\boldsymbol \alpha$ respectively acting in the chiral (valley) and  spin spaces. The spinor is defined as
$\Psi^{\top}_{\mathbf{p}}= \left[ c_{\mathbf{p},\uparrow,+}, c_{\mathbf{p},\downarrow,+}, c_{\mathbf{p},\uparrow,-}, c_{\mathbf{p},\downarrow,-}\right]$, where $c_{\mathbf{p},\sigma,\tau}$ is the fermion annihilation operator with momentum $\mathbf{p}$ (measured from the Weyl nodes), spin projection $\alpha=\uparrow/ \downarrow$, and chirality $\tau=+/-$ (left/right). As shown in the Supplementary Information (SI), the above low-energy Hamiltonian for WSM can be realized from a simple tight-binding model on a cubic lattice. Integrals over momentum run up to an ultraviolet cutoff $\Lambda \sim 1/a$, with $a$ being the lattice spacing. The above Hamiltonian enjoys a global chiral U(1) symmetry, generated by $\gamma_5=\tau_3 \otimes \alpha_0$, which in the continuum limit also stands as the generator of translational symmetry~\cite{roy-sau}.

\emph{Disorder}: Weyl fermions are susceptible to various disorder and the scattering processes by different types of impurities\cite{balatsky}, represented by potential terms coupled to appropriate fermion bilinears. Effects of randomness are captured by the Euclidean action $S_D= \int d^3 x d\tau \; V_N(\mathbf{x})$ $\left( \Psi^\dagger \hat{N} \Psi \right)$, where $V_N(\mathbf{x})$ for simplicity assumes a Gaussian white noise distribution, with disorder average $\langle \langle V_N(\mathbf{x}) V_N(\mathbf{x}^\prime) \rangle \rangle =\Delta_N \delta(\mathbf{x}-\mathbf{x}^\prime)$. As shown in the SI, various types of disorder can be described by an appropriate choice of the $4 \times 4$ matrix $\hat{N}$ and the scaling dimension of disorder coupling $[\Delta_N]=2 z-d$. The clean WSM features linearly dispersing quasiparticles and is thus characterized by dynamic scaling exponent $z=1$. Therefore, sufficiently weak disorder is an \emph{irrelevant} perturbation, since $[\Delta_N]=-1$, and low energy excitations retain their ballistic nature in weakly disordered WSMs.

 The fact that weak disorder is an irrelevant perturbation in three-dimensional WSM gives rise to the possibility of a disorder-driven QPT to a metallic phase for strong  disorder~\cite{fradkin, shindou, ominato, chakravarty, roy-dassarma, radzihovsky, altland, imura, herbut-disorder, brouwer-1, pixley-1, brouwer-2, pixley-2, ohtsuki, roy-bera, hughes}. In light of this, we next show that OC  exhibits a single parameter scaling, and therefore can serve as a bonafide OP across the WSM-metal QPT. For extremely strong disorder, the three-dimensional metal eventually undergoes the \emph{Anderson transition} into an insulating phase~\cite{fradkin, pixley-1}, which is, however, outside the scope of the current work.

\emph{Scaling}: The scaling of conductivity ($\sigma$) with the system size ($L$) follows from the gauge invariance, leading to $\sigma \sim L^{2-d}$, see SI. As the system approaches the QCP located at $\Delta_N=\Delta^{\ast}_N$, the correlation length ($\xi$) diverges according to $\xi \sim |\delta|^{-\nu}$, while the corresponding energy ($\epsilon_0$) vanishes as $\epsilon_0 \sim |\delta|^{\nu z}$, where $\delta=(\Delta_N-\Delta^{\ast}_N)/\Delta^{\ast}_N$ measures the distance from the QCP. Therefore, semimetallic and metallic phases are respectively realized for $\Delta_N<\Delta^{\ast}_N$ and $\Delta_N>\Delta^{\ast}_N$. In the proximity to a QCP, the universal scaling of any physical observable depends on two dimensionless parameters $L/\xi$ and $\Omega/\epsilon_0$. Thus general scaling theory and gauge invariance dictate the following scaling ansatz for the OC (in units of $e^2/h$) in a dirty WSM
\begin{align}\label{Eq:OC}
\sigma(\Omega,\delta,L) = L^{2-d} {\mathcal G} \left( \frac{L}{\delta^{-\nu}}, \frac{\Omega}{\delta^{\nu z}}\right)= \delta^{\nu (d-2)} {\mathcal F} \left( \frac{L}{\delta^{-\nu}}, \frac{\Omega}{\delta^{\nu z}} \right),
\end{align}
 where ${\mathcal G}$ and ${\mathcal F}$ are two unknown, but universal scaling functions. Although the explicit forms for these scaling functions are generally unknown and can only be determined experimentally, their salient features can be deduced from the behavior of OC in various phases of a dirty WSM. Since we are interested in the optical properties of a WSM in the thermodynamic limit $(L \to \infty)$, for brevity we drop the explicit $L$-dependence in $\sigma(\Omega,\delta, L)$. Although we here exploit the gauge invariance and scaling theory to obtain the scaling ansatz in Eq.~(\ref{Eq:OC}), this can also be achieved from the renormalization group analysis of the disorder coupling, as shown in the SI.  When the Fermi energy ($E_F$) is pinned away from the Weyl points (see red dashed line in Fig.~\ref{criticalregime}), the system behaves as a diffusive metal at the lowest energy scale for arbitrary strength of impurity scatterers and our discussion on the scaling of OC is germane only for $\Omega > E_F$.

First we focus on the QCP ($\delta=0$), where the OC must be devoid of any $\delta$-dependence, dictating ${\mathcal F}(x) \sim x^{(d-2)/z}$.  Its scaling  with frequency is then given by
\begin{equation}\label{OCQCP}
\sigma(\Omega, \delta=0) \equiv \sigma_{Q}(\Omega) \sim \Omega^{(d-2)/z}. \nonumber
\end{equation}
Therefore, within the critical regime OC vanishes with a peculiar power-law dependence when frequency $\Omega > \epsilon_0 \sim |\delta|^{\nu z}$, which in turn roughly determines the extent of the critical regime at finite frequencies (see Fig.~\ref{criticalregime}). Notice that as the QCP is approached from the WSM phase the residue of quasiparticle pole vanishes smoothly~\cite{roy-dassarma}, while approaching it from the metallic side the diffusion coefficient diverges~\cite{herbut-disorder}. Therefore, the critical regime constitutes a \emph{non-Fermi liquid} phase of strongly coupled gapless critical modes, due to quantum fluctuations driven by disorder, where the OC scales as $\Omega^{1/z}$.

Next we consider the metallic phase, where average density of states at zero energy is finite, and thus OC as $\Omega \to 0$ also becomes finite due to a finite lifetime of diffusive fermions. Hence, inside the metallic phase ${\mathcal F}(x) \sim x^{0}$ (to the leading order) and OC scales as
$$ \sigma(\Omega \to 0, \delta>0) \equiv \sigma_{M} (\Omega \to 0) \sim \delta^{\nu(d-2)}. $$
OC in the metallic phase thus depends only on $\nu$ as $\Omega \to 0$, which together with the dependence of $\sigma_{Q}(\Omega)$ solely on $z$ endows a unique opportunity to extract the correlation length and the dynamic critical exponents near the WSM-metal QCP independently, and that way pin the universality class of this transition. Hence, in the presence of strong disorder ($\Delta_N>\Delta^{\ast}_N$), as the frequency is gradually lowered the intra-band component of OC starts to dominate over the inter-band counterpart, and in the limit $\Omega \to 0$, only the former contribution survives. Therefore, in the super-critical regime, OC displays a smooth crossover from $\Omega^{1/z}$ dependence (high frequency) toward a constant value as $\Omega \to 0$ (low frequency) around $\Omega \sim \delta^{\nu z}$ [see Fig.~\ref{criticalregime}].  The Drude-peak (arising from the intraband contribution) inside the metallic phase gets broadened due to a finite transport lifetime of quasiparticles, and its width increases with the strength of disorder. By contrast, inside the WSM phase and quantum critical regime the Drude-peak remains sharp.

Finally, we delve into the scaling of OC on the WSM side of the transition. In the clean limit, on dimensional grounds, we expect inter-band OC $\sigma(\Omega) \sim \Omega^{d-2}$. Such scaling of OC remains valid in the weakly disordered WSM, at least when $\Omega \ll \epsilon_0$, indicating that ${\mathcal F}(x) \sim x^{d-2}$ for $\delta<0$, leading to
\begin{equation}\label{OCWSM}
\sigma(\Omega, \delta<0) \equiv \sigma_{W} (\Omega) \sim \Omega^{d-2} \; |\delta|^{\nu (1-z)(d-2)}, \nonumber
\end{equation}
which vanishes linearly with frequency $\Omega$. With increasing strength of disorder, the system becomes more metallic and typically at the WSM-metal QCP $z>1$~\cite{fradkin, chakravarty, roy-dassarma, radzihovsky, herbut-disorder, pixley-1, brouwer-2, pixley-2, ohtsuki, roy-bera}. Consequently, as one approaches the WSM-metal QCP from the semimetallic side, the system becomes more metallic and $\sigma_{W} (\Omega)$ increases. In the weak disorder regime, the inter-band component of OC dominates over intra-band piece until $\Omega \sim E_F$, with $E_F$ being the Fermi energy (typically unknown) in a WSM, and with increasing frequency OC displays a smooth crossover from $\Omega$ to $\Omega^{1/z}$ dependence. As disorder increases the frequency range over which OC scales linearly with the frequency shrinks, while the region with $\Omega^{1/z}$ scaling increases. Finally, at the WSM-metal QCP $\sigma \sim \Omega^{1/z}$ over the entire range of frequency [see Fig.~\ref{criticalregime}], at least when $\Omega \ll 2 v \Lambda$.

\emph{Optical response in a WSM}: Since weak disorder flows toward smaller values with increasing RG time $l \sim \log(v\Lambda/\Omega)$
or decreasing frequency, the lowest energy excitations are described by ballistic chiral fermions in a weakly disordered WSM. Thus, we can rely on the \emph{Kubo formalism} in this regime to compute OC of a WSM diagramatically, and directly test the validity of its scaling ansatz for weak enough randomness. To set the stage, we first focus on the OC in a clean WSM ($\Delta_N=0$), which at zero temperature can be extracted from the current-current correlation function. In what follows and as shown in the SI, we compute the integrals over the internal momentum in $d=3-\epsilon$ spatial dimensions, and at the end send $\epsilon \to 0$, closely following the spirit of \emph{dimensional regularization} that manifestly preserves the \emph{gauge invariance}~\cite{juricic, HJV2008}. The OC in a clean WSM is $\sigma( \Omega ) = (e^2 N_f \Omega)/( 6 h v) \equiv \sigma_0$,
with $N_f$ as the number of Weyl pairs. In this limit ($\Delta_N=0$), $\sigma_{W} (\Omega)=\sigma(\Omega)$, in agreement with the above scaling form. Therefore, inter-band component of OC scales \emph{linearly} with the frequency~\cite{chakravarty, hosur, rosenstein,vieri}, as has been observed in Nd$_2$(Ir$_x$Rh$_{1-x}$)$_2$O$_7$~\cite{ueda} and Eu$_2$Ir$_2$O$_7$~\cite{drew}, which possibly through an ``all-in all-out" magnetic ordering in pyrochlore lattice enter into a WSM phase~\cite{vishwanath}.

By now it is well established that random charge impurities ($\Delta_V$) can drive WSM-metal transition~\cite{fradkin, shindou, ominato, chakravarty, roy-dassarma, radzihovsky, altland, imura, herbut-disorder, brouwer-1, pixley-1, brouwer-2, pixley-2, ohtsuki, roy-bera, hughes} or at least can support a large crossover regime if rare regions dominate at the lowest energy scale~\cite{pixley-rareregion}. As described in depth in the SI, elastic scatterer of any other nature (magnetic, spin-orbit, mass disorder, etc.) generates random \emph{axial chemical potential} through quantum corrections. The axial disorder ($\Delta_A$) causes random but equal and opposite shifts of the Fermi level for left and right chiral fermions, while maintaining the overall charge neutrality of the system. Strong axial disorder also gives rise to semimetal-metal QPT~\cite{chakravarty, roy-dassarma, pixley-1, pixley-2}. Hence, to anchor the scaling behavior of OC in weak disorder regime, it is sufficient to focus on these two disorder couplings, $\Delta_V$ and $\Delta_A$, respectively characterized by two matrices $\hat{N}=\tau_0 \otimes \alpha_0$ and $\hat{N}=\tau_3 \otimes \alpha_0$. After accounting for the correction to OC to the lowest order in disorder coupling the total OC is given by (see SI for details)
\begin{equation}\label{OC-discorrected}
\sigma(\Omega)= \sigma_0 \; \left[ 1+ \; \hat{\Delta}_N \; F \left(\frac{\Omega}{2 \Lambda v}\right) \right],
\end{equation}
for $N=V,A$, where $\hat{\Delta}_N= \Delta_N \Lambda/(\pi^2 v^2)$ is the dimensionless bare disorder strength, and the function $F(x) \approx 1-x^2+ {\mathcal O} (x^4)$. The above form of the OC $\sigma(\Omega)=\sigma_0 (1+\hat{\Delta}_N)$ is also compatible with the scaling form of $\sigma_{W} (\Omega)$ after substituting $z=3/2$, $\nu=1$, as predicted from one-loop RG calculation~\cite{chakravarty, roy-dassarma, radzihovsky} and also reasonably consistent with recent numerical works~\cite{herbut-disorder, pixley-2, ohtsuki, roy-bera}, with $\hat{\Delta}^\ast_{V/A} =1/2$ being the non-universal critical strength of disorder for the WSM-metal QPT. Such a striking agreement between scaling theory [see Eq.~(\ref{Eq:OC}) and $\sigma_W(\Omega)$], perturbative correction to OC in the weak disorder limit [see Eq.~(\ref{OC-discorrected})], RG and numerical analyses indicates internal consistency of our analysis, and puts forward OC as a bonafide OP across the unconventional QPT from  WSM to a metallic phase. With one-loop result for the critical exponents $\nu$ and $z$, OC in the critical regime $\sigma_Q(\Omega) \sim \Omega^{2/3}$ and inside the metallic phase $\sigma_M(\Omega \to 0) \sim \delta$. However, as our scaling analysis suggests, these critical exponents  can be determined independently from the scaling of OC in numerical studies and experiments to precisely determine the universality class of the WSM-metal transition. Furthermore, as shown in the SI, scaling of OC as $\Omega \to 0$ with system size ($L$) inside the metallic phase allows one to extract the correlation length exponent ($\nu$) independently.

The imaginary part of OC in a weakly disordered WSM also receives a correction  yielding  the total dielectric constant in the presence of the chemical potential ($N=V$) or axial disorder ($N=A$)
\begin{equation}\label{dielectric}
\varepsilon(\Omega)=1+\frac{2 e^2 N_f}{3 h v} \left\{ \log \left[ \frac{4 v^2 \Lambda^2}{\Omega^2} -1\right]- \frac{\Omega \Delta_N}{4 v^3} \right\},
\end{equation}
which displays a \emph{logarithmic} enhancement as $\Omega \to 0$. It is worth mentioning that recent experiment has observed enhancement of $\varepsilon (\Omega)$ in Eu$_2$Ir$_2$O$_7$ as $\Omega \to 0$~\cite{drew}.

Furthermore, in the presence of arbitrary disorder the OC in three-dimensional WSM exhibits a remarkably universal dependence on frequency and disorder, but up to a sign, depending on the type of elastic scatterer, as shown in Table~I of the SI. Correction to the OC due to any disorder (such as the spin-orbit one with $\Delta_N=\Delta_{SO}$ and $\hat{N}=\tau_2 \otimes \alpha_j$, where $j=1,2,3$) that together with the axial disorder also drives a WSM-metal QPT through a QCP that, however, belongs to a different universality class (with $\nu=1$, but $z=11/2$ to one-loop order~\cite{chakravarty, roy-intdis}), also conforms to the critical scaling form shown in Eq.~(\ref{OC-discorrected}). Furthermore, in the presence of both potential and axial disorders, WSM-metal QPT takes place through a line of QCPs in the $(\Delta_V, \Delta_A)$ plane along which $\nu=1$ and $z=3/2$ (to one-loop order)~\cite{chakravarty, roy-intdis}. The OC then reads as $\sigma(\Omega)=\sigma_0 \left( 1+ \hat{\Delta}_V + \hat{\Delta}_A\right)$, which also conforms to the universal scaling form of the OC, since the line of QCPs is given by $\hat{\Delta}^\ast_V + \hat{\Delta}^\ast_A=1/2$. Finally, the dielectric constant also receives a universal (up to a sign) correction due to disorder that scales linearly with frequency, as shown in Eq.~(\ref{dielectric}).

\emph{Discussion}: We establish OC as an experimentally accessible OP across the disorder-driven WSM-metal QPT.  In particular, we show that it can uncover signatures of an underlying dirty QCP by exposing the associated quantum critical regime at finite frequencies. While the scaling analysis is performed here strictly at $T=0$ in the ballistic (collisionless) regime, it remains operative also at finite temperature as long as $\Omega\gg T$~\cite{wegner}. The finite conductivity in the metallic phase as $\Omega \to 0$ for stronger disorder should match the dc conductivity when $T \neq 0$\cite{huang-dassarma1,huang-dassarma2}; the value of the former is, however, expected to be different if temperature is set to be zero first. Nevertheless, irrespective of these two limits $\sigma_M (\Omega \to 0)$ follows the announced scaling behavior. Our scaling arguments can also be applied to the dc conductivity~\cite{chakravarty, radzihovsky, brouwer-1} in the \emph{collision-dominated} regime $T \gg \Omega$, for which the scaling behavior qualitatively follows Eq.~(\ref{Eq:OC}) upon taking $\Omega \to T$~\cite{wegner}.

Even though we primarily focused on WSMs~\cite{taas-1, tasas-2, taas-3,nbas-1,tap-1,nbp-1, nbp-2,tas,borisenko, chiorescu}, our results are consequential to a vast number of materials, such as the topological Dirac semimetal that has recently been discovered in Cd$_2$As$_3$~\cite{cdas} and Na$_3$Bi~\cite{nabi}, conventional Dirac semimetals that can be found at the QCP separating two topologically distinct ( for example, strong, weak, crystalline and trivial) insulating phases in various three-dimensional strong spin-orbit coupled materials, such as Bi$_{1-x}$Sb$_{x}$, Bi$_2$Se$_3$, Bi$_2$Te$_2$, Sb$_2$Te$_3$~\cite{TI-review-1,TI-review-2} and quasi-crystals supporting Dirac fermions~\cite{basov}. From the extent of the critical regime and semimetalic phase at finite frequencies (see Fig.~\ref{criticalregime}), we expect that the critical scaling of OC and its correction due to random impurities can be observed in a broad class of disordered Weyl and Dirac semimetals.

%

\section*{Acknowledgements}
B. R. and S. D. S. are supported by NSF-JQI-PFC and LPS-MPO-CMTC. B. R. is thankful to H. Dennis Drew, Pallab Goswami and Igor Herbut for stimulating discussions, and Aspen Center for Physics for hospitality during  the Summer Program (2015), where part of this work was finalized.

\section*{Author contributions statement}

All authors contributed equally to the conceptual development of the problem and participated in writing the manuscript. B. R. and V. J. performed the calculations.

\section*{Additional information}

\textbf{Competing financial interests:} The authors declare no competing financial interests.


\pagebreak

\onecolumngrid

\begin{center}
{\bf Supplementary Materials for ``\emph{Universal optical conductivity of a disordered Weyl semimetal}"} \\

Bitan Roy$^{1}$, Vladimir Juri\v ci\' c$^{2}$, Sankar Das Sarma$^{1}$ \\

$^{1}$\emph{Condensed Matter Theory Center and Joint Quantum Institute, University of Maryland, College Park, Maryland 20742-4111, USA} \\

$^{2}$\emph{Nordita,  Center for Quantum Materials,  KTH Royal Institute of Technology and Stockholm University, Roslagstullsbacken 23,  10691 Stockholm, Sweden}
\end{center}

This Supplementary Information contains:
\begin{enumerate}
\item{A lattice model for Weyl fermions.}
\item{The scaling analysis for disorder coupling in three dimensional Weyl semimetals.}
\item{Generation of axial disorder from arbitrary disorder through quantum loop corrections.}
\item{Derivation of scaling dimension of optical conductivity.}
\item{Calculation of optical conductivity and dielectric constant in a clean Weyl semimetal.}
\item{Corrections to optical conductivity and dielectric constant due to weak randomness in the system.}
\item{Some essential integral identities useful for the calculation of optical conductivity.}
\item{Derivation of the scaling of optical conductivity from renormalization group flow equations.}
\item{Universal scaling of finite metallic conductivity (as frequency $\Omega \to 0$) with system size ($L$).} 
\end{enumerate}

$\bullet$ \emph{\bf A lattice model of Weyl semimetal.} The low energy Hamiltonian for WSM, shown in the main part of the paper, can be realized from a simple tight-binding model on a cubic lattice
\begin{eqnarray}\label{weyllattice}
H_W = t \; \sum_{\mathbf{k}} \Psi^\dagger_{\mathbf{k}} \; \big[ \sigma_1 \cos(k_1 a) + \sigma_2 \cos(k_2 a) + \sigma_3 \cos(k_3 a)
+ \frac{t^\prime}{t} \: \sigma_{3} \left(\sin(k_1 a)+\sin(k_2 a)-2 \sin(k_3 a) \right) \big] \Psi_{\mathbf{k}},
\end{eqnarray}
where $a$ is the lattice spacing and $\Psi^{\top}_{\mathbf{k}}=[c_{\mathbf{k},\uparrow}, c_{\mathbf{k},\downarrow}]$ is a two component spinor, and $c_{\mathbf{k},\sigma}$ is the fermion annihilation operator with momentum $\mathbf{p}$ and spin projection $\alpha=\uparrow/ \downarrow$. The term proportional to $t^\prime$ plays the role of momentum dependent \emph{Wilson mass} that turns three pairs of Weyl nodes into massive, except the ones at $\pm \mathbf{k}_0$, respectively supporting left and right chiral fermions, where $\mathbf{k}_0= \left( 1,1,1\right) \frac{\pi}{2 a}$. The low energy Hamiltonian in Eq.~(1) of main paper is then obtained by linearizing the above tight binding model near $\pm \mathbf{k}_0$, with $v=t a/\hbar$, $\Lambda \sim 1/a$ is the ultraviolet cutoff, and momentum ($\mathbf{p}$) is measured from two Weyl nodes.
\\

$\bullet$ \emph{\bf Scaling dimension of disorder in WSM.} The effects of disorder can be captured from the Euclidean action $S_D= \int d^3 x d\tau \; V_N(\mathbf{x})$ $\left( \Psi^\dagger \hat{N} \Psi \right)$. We assume that all disorder assumes Gaussian white noise distribution, with disorder average $\langle \langle V_N(\mathbf{x}) V_N(\mathbf{x}^\prime) \rangle \rangle =\Delta_N \delta(\mathbf{x}-\mathbf{x}^\prime)$. Then performing the disorder averaging, we arrive at the replicated action
\begin{eqnarray}\label{action}
\bar{S}=\int d^d \vec{x} d \tau \; \left(\bar{\Psi}_{a} \left[-i \partial_\tau \gamma_0-i \hbar v \; \gamma_j \partial_j \right] \Psi_{a}\right)_{(\tau,\vec{x})}
-\frac{\Delta_N}{2} \int d^d \vec{x} d \tau d \tau^\prime \left(\bar{\Psi}_{a} N \Psi_{a} \right)_{(\vec{x},\tau)} \left(\bar{\Psi}_{b} N \Psi_{b} \right)_{(\vec{x},\tau^\prime)},
\end{eqnarray}
where $a,b$ are replica indices, and $\bar{\Psi}=\Psi^\dagger \gamma_0$ as usual. For convenience we have slightly modified the definition of the $\gamma$ matrices from the main part of the paper according to $\gamma_0=\tau_1 \otimes \alpha_0$, $\gamma_j=\tau_2 \otimes \alpha_j$ for $j=1,2,3$ and $\gamma_5=\tau_3 \otimes \alpha_0$. These five matrices constitutes the Clifford algebra of maximal number of mutually anticommuting four dimensional matrices. Various types of elastic scatters (disorder) are realized with different choices of $4 \times 4$ matrices, as shown in Tab.~\ref{table-disorder}. Notice that due to linearly dispersing quasiparticles WSM corresponds to a $z=1$ fixed point in $d=3$. Consequently under the rescaling of space-time(imaginary) coordinates $\left( \mathbf{x}, \tau \right) \to e^{l}\left( \mathbf{x}, \tau \right)$. The Euclidean action $\bar{S}$ in Eq.~(\ref{action}) remains invariant under such coarse graining only when accompanied by rescaling of fermionic field according to $\Psi \to e^{-dl/2} \Psi$. The scaling dimension of disorder coupling then reads as $[\Delta_N]=2z-d=-1$ for $d=3$, for any choice of $N$. Therefore, any weak disorder is an irrelevant perturbation at the $z=1$ fixed point, describing a clean WSM.
\\

\begin{figure}[htb]
\subfigure[]{
\includegraphics[width=5.5cm, height=2.5cm]{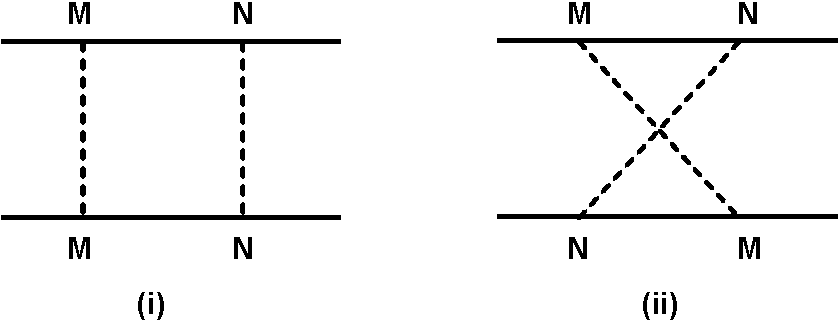}
\label{laddercrossing}
}
\subfigure[]{
\includegraphics[width=7.0cm, height=2.25cm]{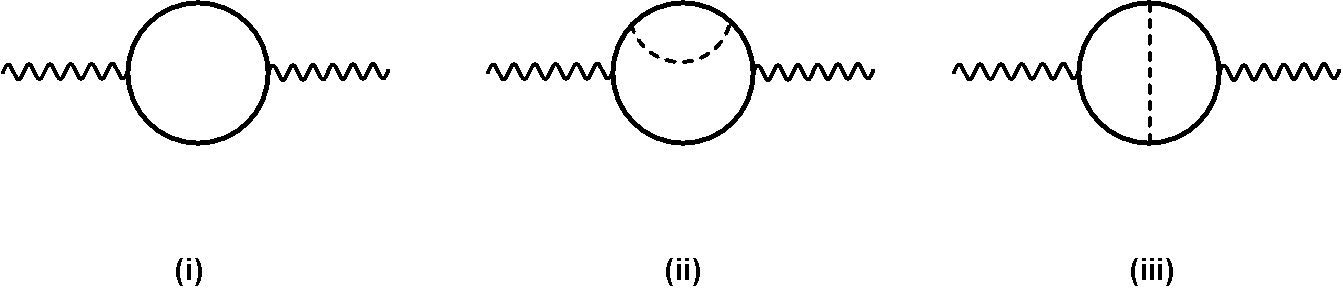}
\label{conductivity}
}
\subfigure[]{
\includegraphics[width=4.0cm, height=2.25cm]{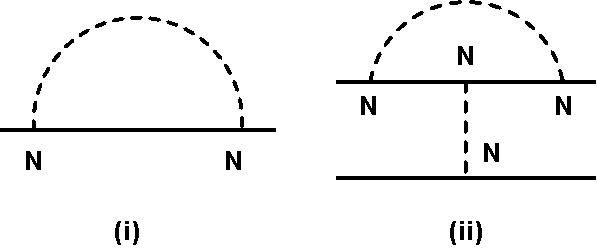}
\label{selfenergy-vertex}
}
\caption[] { (a) One-loop diagrams that can generate axial disorder ($\Delta_A$), here $M$ and $N$ are two $4 \times 4$ matrices. (b) OC in clean WSM arises from diagram (i), while its leading order correction due to weak disorder stems from diagrams (ii) and (iii). (c) Relevant one-loop diagrams to capture the effects of potential/axial disorder with $N=\gamma_0/\gamma_0\gamma_5$. For $M=N$, two diagrams in (a) cancel each other out. Solid, dashed and wavy lines respectively represent fermions, disorder and external gauge field. }\label{feynman-conductivity}
\end{figure}

$\bullet$ \emph{\bf Generation of axial disorder in renormalization group framework.} We here show how axial disorder $(\Delta_A)$ can be generated from all types of disorder (except the potential disorder) once we account for quantum (loop) corrections. For this purpose computation of simple one-loop diagrams, shown in Fig.~\ref{laddercrossing}, is sufficient and serves the purpose. Total contribution from these two diagrams reads as
\begin{eqnarray}
(\mathrm{a, i})+(\mathrm{a, ii})= \Delta_N \Delta_M \; \int^{\prime} \frac{d^{3} \mathbf k}{(2 \pi)^3} \: \left( \bar{\Psi}_a M G_0(0, \mathbf k) N\Psi_a\right) \: \left\{ \bar{\Psi}_b \left( N G_0(0, \mathbf k) M + M G_0(0, -\mathbf k) N \right) \Psi_b \right\},
\end{eqnarray}
where the integral over momentum is carried out within the Wilsonian shell $\Lambda e^{-l}< |{\mathbf k}|<\Lambda$ and $\Delta_{N/M}$ captures the strength of different types of elastic scatterer characterized by $4 \times 4$ matrix $N/M$ (see Table~\ref{table-disorder}). The noninteracting Green's function is given by
\begin{equation}
G_0(i\omega, {\bf k})=- \; \frac{i\gamma_0\omega+i\gamma_j k_j}{\omega^2+k^2}.
\end{equation}
The total contribution from these two diagrams from magnetic, spin-orbit, current and axial-magnetic impurities reads as
\begin{eqnarray}\label{axialgeneration}
(\mathrm{a, i})+(\mathrm{a, ii})= \left(\; \Delta^2_M + \Delta^2_{AM} +\Delta^2_{SO}+\Delta^2_C \; \right) \; \left( \frac{\Lambda \; l}{\pi^2 v^2} \right) \; \left( \bar{\Psi}_a \gamma_0 \gamma_5 \Psi_a \right) \; \left( \bar{\Psi}_b \gamma_0 \gamma_5 \Psi_b \right),
\end{eqnarray}
where $a$ and $b$ are replica indices. Thus aforementioned four disorders generate axial disorder.
\\

The effect of mass disorder in WSM is more subtle. In Weyl semimetals the scalar ($\Delta_S$) and pseudo-scalar ($\Delta_{PS}$) mass disorder are present simultaneously. The quantum corrections arising from diagrams (i) and (ii) of Fig.~\ref{laddercrossing} in the presence of two mass disorder $\Delta_{S}$ and $\Delta_{PS}$ yield
\begin{equation}
(\mathrm{a, i})+(\mathrm{a, ii})= \frac{\Delta_S \Delta_{PS}}{3} \; \left( \frac{\Lambda \; l}{\pi^2 v^2} \right) \; \left( \bar{\Psi}_a i \gamma_5 \gamma_j \Psi_a \right) \; \left( \bar{\Psi}_b i \gamma_5 \gamma_j \Psi_b \right),
\end{equation}
where $j=1,2,3$. Therefore, mass disorder ($\Delta_{S}$ and $\Delta_{PS}$) in Weyl semimetal generates magnetic disorder ($\Delta_{M}$) through loop corrections. As we have shown in Eq.~(\ref{axialgeneration}),  magnetic disorder in turn generates axial disorder. Therefore, to appreciate the scaling of optical conductivity it is sufficient to examine the effects of regular charge impurity ($\Delta_V$) and axial disorder ($\Delta_A$). A complete renormaliazation group analysis in the presence of all disorder couplings is an interesting problem, which however, falls outside the central theme of current work and deserves a separate publication~\cite{roy-juricic-unpublished}. Nevertheless, the above exercise justifies our claim that salient features of the scaling properties of optical conductivity can be captured by focusing on (1) random charge impurity and (2) random axial chemical potential.
\\

\begin{table}[h]
\begin{tabular}{|l|l|l|l|l|l|l|l|}
\hline
Bilinear & Physical quantity & $\mathcal{T}$ & $\mathcal{P}$ & $U_c$ & $\mathcal{C}$ & disorder coupling & $f(\Delta_N)$ \\
\hline \hline
$\bar{\Psi} \gamma_0 \Psi$ & chemical potential   &  $\checkmark$ & $\checkmark$  & $\checkmark$ &  $\times$ & $\Delta_V$ & $+$    \\
\hline
$\bar{\Psi} \gamma_0 \gamma_5 \Psi$ & axial potential  & $\checkmark$  & $\times$ & $\checkmark$ &  $\checkmark$ & $\Delta_A$ & $+$ \\
\hline
$\bar{\Psi}\Psi$ & scalar mass & $\times$  & $\checkmark$  & $\times$ & $\checkmark$ & $\Delta_S$ & $-$ \\
\hline
$\bar{\Psi}i \gamma_5\Psi$ & pseudo-scalar mass & $\checkmark$ & $\times$  & $\times$  & $\checkmark$ & $\Delta_{PS}$ & $-$\\
\hline
$\bar{\Psi}i \gamma_5 \gamma_j \Psi$ & magnetization & $\times$  & $\checkmark$  & $\checkmark$ & $\checkmark$ & $\Delta_M$ & $-$ \\
\hline
$\bar{\Psi}i \gamma_j \Psi$ & current & $\times$  & $\times$  & $\checkmark$ & $\times$ & $\Delta_C$ & $-$  \\
\hline
$\bar{\Psi}\gamma_0 \gamma_j \Psi$ & spin-orbit coupling & $\times$  & $\times$  & $\times$ & $\times$ & $\Delta_{SO}$ & $+$ \\
\hline
$\bar{\Psi} i \gamma_0 \gamma_5 \gamma_j \Psi$ & axial-magnetization & $\checkmark$  & $\checkmark$  & $\times$ & $\times$ & $\Delta_{AM}$ & $+$ \\
\hline
\end{tabular}
\caption{ Various types of disorder represented by  fermionic bilinears ($j=1,2,3$), their symmetries under pseudo time-reversal ($\mathcal{T}$), parity ($\mathcal{P}$), continuous chiral rotation ($U_c$) and charge-conjugation ($\mathcal{C}$). The disorder couplings are represented by $\Delta_N$, and the corresponding sign function is given by $f(\Delta_N)$. Note that true time-reversal symmetry in WSM in already broken. The pseudo time-reversal symmetry ${\mathcal T}$ is generated by an anti-unitary operator $\gamma_0 \gamma_2 K$, where $K$ is complex conjugation, such that ${\mathcal T}^2=-1$ (The true time-reversal operator is $\gamma_1 \gamma_3 K$). The parity operator is generated ${\mathcal P}=\gamma_0$, while the charge-conjugation operator is ${\mathcal C}=\gamma_2$. The continuous chiral symmetry ($U_c$) of WSM is generated by $\gamma_5$, the generator of translational symmetry in the continuum limit in clean Weyl semimetal~\cite{roy-sau}. Here $\checkmark$ and $\times$ signifies even and odd under a symmetry operation, respectively. }\label{table-disorder}
\end{table}

$\bullet$ \emph{\bf Derivation of scaling dimension of optical conductivity.} The real part of optical conductivity at zero momentum is given in terms of the analytically continued current-current correlation function as
\begin{equation}\label{eq:real-part-optical cond}
\Re\left(\sigma_{lm}\right) = \lim_{\delta \to 0}\frac{\text{Im}\; \Pi_{lm}(i \Omega \to \Omega+i \delta)}{\Omega},
\end{equation}
while the imaginary part is
\begin{equation}
\Im (\sigma_{lm})= -\lim_{\delta \to 0}\frac{\text{Re} \; \Pi_{lm}(i\Omega \to \Omega+i \delta)}{\Omega}.
\end{equation}
 The current-current correlation function is
\begin{equation}\label{eq:polarization}
\Pi_{lm}(i\Omega)=\int d\tau\; d^d \vec{x}\;e^{-\Omega\tau}\langle j_l(\vec{x},\tau)j_m(0,0)\rangle=\Pi(i \Omega)\delta_{lm},
\end{equation}
with the last equality valid in isotropic systems, which we exclusively consider here.

To find the scaling dimension of the optical conductivity, we first need to find the scaling dimension of the current. The latter is obtained from the coupling of  the external vector potential to the fermions by the minimal substitution $\partial_j\rightarrow\partial_j-ie A_j$. Consequently, the gauge invariant current-gauge field coupling is
\begin{equation}\label{eq:j-A coupling}
S_{j-A}=\int d^d\vec{x}\; d \tau \; j_i A_i.
\end{equation}
The scaling dimension of the gauge field is $[A]=-1$ in units of the length, i.e. $[L]=1$, and therefore $[j]=-z-d+1$. Here, $z$ is the dynamical critical exponent which determines the relative scaling between the frequency and the momentum, $\omega\sim k^z$, and therefore $[\omega]=-z$. From Eq.\ (\ref{eq:polarization}), we then obtain $[\Pi]=+d+z+2[j]=-z-d+2$. Finally, from Eq.\ (\ref{eq:real-part-optical cond}) it follows that the scaling dimension of the optical conductivity is $[\sigma]=-z-d+2+z=2-d$. The scaling dimension of the optical conductivity is protected by the gauge invariance and therefore does not receive any correction, i.e. its anomalous dimension is precisely equal to zero.\\

$\bullet$ \emph{\bf Optical properties of a clean WSM.} We now present the calculation of the optical conductivity (OC) of a  clean WSM. The OC in clean WSM at zero temperature ($T=0$) can be extracted from the current-current correlation function (see Fig.~\ref{conductivity}(i) for corresponding Feynman diagram), which as a function of external frequency (Matsubara) has the form
\begin{equation}
\Pi^{(0)}_{lm}(i\Omega)=\frac{e^2}{h}\frac{1}{v^{d-2}}{\rm Tr}\int\frac{d\omega d^d{\bf k}}{(2\pi)^d}\gamma_l G_0(i(\omega+\Omega),{\bf k})\gamma_m
G_0(i\omega,{\bf k}),
\end{equation}
where ${\rm Tr}$ denotes the trace over the $\gamma$-matrices. Performing the above trace and  integral in $d=3$, we obtain
\begin{equation}\label{pinonint}
\Pi_{lm}^{(0)}(i \Omega)= -\left( \frac{e^2}{h}\right)\:\frac{N_f \; \Omega^2}{6 \pi v} \log \left[1+ \frac{4 v^2 \Lambda^2}{\Omega^2} \right]\;\delta_{lm},
\end{equation}
where $N_f$ is the number of Weyl pairs. After performing the \emph{analytic continuation} to real frequency $i \Omega \to \Omega +i \delta$ (Kubo formalism), we obtain the real part of OC
\begin{equation}
\Re\left(\sigma\right) = \lim_{\delta \to 0}\frac{\text{Im}\; \Pi(i \Omega \to \Omega+i \delta)}{\Omega} = \frac{e^2 N_f \; \Omega}{ 6 h v} \equiv \sigma_0.
\end{equation}
 Therefore, inter-band component of OC vanishes \emph{linearly} with the frequency ($\Omega$), and a clean three-dimensional WSM in strict sense is a \emph{power-law insulator}.
\\

The imaginary part of the OC is given by
\begin{equation}
\Im (\sigma)= -\lim_{\delta \to 0}\frac{\text{Re} \; \Pi(i\Omega \to \Omega+i \delta)}{\Omega}= -\frac{\sigma_0}{\pi} \log \left[ \frac{4 v^2 \Lambda^2}{\Omega^2}-1\right],
\end{equation}
which reconciles with the result obtained by using the Kramers-Kronig (KK) relation to the real part of OC
\begin{eqnarray}
\Im (\sigma) &=& -\frac{2 \Omega}{\pi} \; \mathcal{P} \; \int^{2 v \Lambda}_{0} d \Omega^\prime \: \frac{ \Re\left(\sigma\right)}{{\Omega^\prime}^2-\Omega^2},
\end{eqnarray}
 where $\mathcal{P}$ represents the principal value of the integral. The integral over frequency is limited by the bandwidth of a WSM $2 v \Lambda$. The imaginary part of the OC is related with the real part of dielectric constant according to $\Re(\varepsilon)=1-4 \pi \Im(\sigma)/\Omega$. Thus in a clean WSM, the dielectric function scales as
\begin{equation}
\Re(\varepsilon)=1+ \frac{e^2}{h} \: \frac{2 N_f}{3 v} \: \log\left[ \frac{4 v^2 \Lambda^2}{\Omega^2}-1\right],
\end{equation}
which displays a \emph{logarithmic enhancement} as frequency $\Omega\to 0$.
\\

$\bullet$ \emph{\bf Optical conductivity of weakly disordered WSM.} We now present details of calculation for the correction to the OC in weakly disordered WSM arising from the disorder represented by the $4 \times 4$ matrix $\hat{N}$ in Table \ref{table-disorder}. Corrections to current-current correlation function due to disorder stem from Feynman diagrams (ii) and (iii) in Fig.~\ref{conductivity}, and the total contribution from these two diagrams can be written as
 \begin{equation}
 \delta\Pi^{\hat{N}}_{lm}(i\Omega)=\Sigma^{\hat{N}}_{lm}(i\Omega)+V^{\hat{N}}_{lm}(i\Omega).
 \end{equation}
 The contribution from the self-energy and vertex diagrams in Fig.~\ref{conductivity} (ii) and (iii) are respectively given by
\begin{eqnarray}\label{self-energy}
\Sigma^{\hat{N}}_{lm}(i\Omega)&=&\frac{e^2}{h} \frac{2}{v^{2d-2}}\Delta_N\,{\rm Tr}\int \frac{d^d \mathbf{k}\; d^d \mathbf{p} \; d\omega}{(2 \pi)^{2d}} \:
\gamma_l \, G_0(i(\omega+\Omega),{\bf k})\gamma_m G_0(i\omega,{\bf k})\hat{N} G_0(i\omega,{\bf p}) \hat{N} G_0(i\omega,{\bf k}), \\
V^{\hat{N}}_{lm}(i\Omega)&=&\frac{e^2}{h} \frac{1}{v^{2d-2}} \Delta_N\, {\rm Tr} \int \frac{d^d \mathbf{k}\; d^d \mathbf{p} \; d\omega}{(2 \pi)^{2d}} \:
 \gamma_l \, G_0(i(\omega+\Omega),{\bf k}) \hat{N} G_0(i(\omega+\Omega),{\bf p})\gamma_m G_0(i\omega,{\bf p}) \hat{N} G_0(i\omega,{\bf k}). \label{vertex}
\end{eqnarray}
Different types of disorder correspond to different choices of the matrix $\hat{N}$ in the above two equations [see Table~\ref{table-disorder}].
\\

We fist consider random charge impurity (with $\hat{N}=\gamma_0$) and axial potential disorder (with $\hat{N}=\gamma_0 \gamma_5$) [see Table~\ref{table-disorder}]. For these two types of elastic scatterers contributions from Eqs.~(\ref{self-energy}) and (\ref{vertex}) after performing the trace read as
\begin{eqnarray}\label{self-energy-chemical-potential}
\Sigma^{\hat{N}=\gamma_0 / \gamma_0 \gamma_5}_{lm}(i\Omega)&=& -\delta_{lm} \frac{8 N_f \Delta_N}{v^{2d-2}}\left(\frac{e^2}{h}\right) \int \frac{d^d \mathbf{k}\; d^d \mathbf{p} \; d\omega}{(2 \pi)^{2d}} \:       \frac{\omega(\omega+\Omega)(\omega^2-k^2) - 2(\frac{2}{d}-1) \omega^2 k^2}{((\omega+\Omega)^2+k^2)(\omega^2+k^2)^2 \; (\omega^2+p^2)},  \\
V^{\hat{N}=\gamma_0 / \gamma_0 \gamma_5}_{lm}(i\Omega)&=& -\delta_{lm}\frac{4 N_f \Delta_N}{v^{2d-2}}\left(\frac{e^2}{h}\right) \int \frac{d^d \mathbf{k}\; d^d \mathbf{p} \; d\omega}{(2 \pi)^{2d}} \:
 \frac{\left[\omega(\omega+\Omega)-\left( \frac{2}{d}-1 \right) k^2 \right] \left[\omega(\omega+\Omega)-\left( \frac{2}{d}-1 \right) p^2 \right]}{(\omega^2+k^2)\; (\omega^2+p^2) \; \left[ (\omega+\Omega)^2 +k^2 \right] \; \left[ (\omega+\Omega)^2 +p^2 \right]}, \label{vertex-chemical-potential}
\end{eqnarray}
for $N=V, A$. Upon performing the integrals, we then obtain the correction to the fermion bubble to be
$$\delta\Pi^{\hat{N}=\gamma_0 / \gamma_0 \gamma_5}_{lm}(i \Omega)= \delta_{lm} \: \frac{e^2 N_f\Delta_N}{h} \times \frac{(i\Omega)^3}{24\pi v^4}.$$
\\

Next we account for the scalar and pseudo scalar mass disorder, for which $\hat{N}={\hat 1}_{4 \times 4}$ and $i \gamma_5$, respectively, as shown  in Table~\ref{table-disorder}, where $\hat{1}_{4 \times 4}$ is the $4\times4$ unity matrix.
After performing the trace we obtain
\begin{eqnarray}\label{self-energy-mass-disorder}
\Sigma^{\hat{N}=I_{4\times 4}/i \gamma_5}_{lm}(i\Omega)&=& -\delta_{lm} \frac{8 N_f \Delta_N}{v^{2d-2}}\left(\frac{e^2}{h}\right) \int \frac{d^d \mathbf{k}\; d^d \mathbf{p} \; d\omega}{(2 \pi)^{2d}} \:       \frac{\omega(\omega+\Omega)(\omega^2-k^2) - 2(\frac{2}{d}-1) \omega^2 k^2}{((\omega+\Omega)^2+k^2)(\omega^2+k^2)^2 \; (\omega^2+p^2)},  \\
V^{\hat{N}=I_{4\times 4}/i \gamma_5}_{lm}(i\Omega)&=& \delta_{lm}\frac{4 N_f \Delta_N}{v^{2d-2}}\left(\frac{e^2}{h}\right) \int \frac{d^d \mathbf{k}\; d^d \mathbf{p} \; d\omega}{(2 \pi)^{2d}} \:
 \frac{\left[\omega(\omega+\Omega)-\left( \frac{2}{d}-1 \right) k^2 \right] \left[\omega(\omega+\Omega)-\left( \frac{2}{d}-1 \right) p^2 \right]}{(\omega^2+k^2)\; (\omega^2+p^2) \; \left[ (\omega+\Omega)^2 +k^2 \right] \; \left[ (\omega+\Omega)^2 +p^2 \right]}, \label{vertex-mass-disorder}
\end{eqnarray}
for $N=S, PS$ [see Table~\ref{table-disorder}]. After completing the frequency and momentum integrals we obtain
$$\delta\Pi^{\hat{N}=I_{4\times 4}/i \gamma_5}_{lm}(i \Omega)= -\delta_{lm} \: \frac{e^2 N_f\Delta_N}{h} \times \frac{(i\Omega)^3}{24\pi v^4}.$$
\\

Next we focus on current disorder represented by $\hat{N}=i\gamma_j$ and random magnetic impurities for which $\hat{N}=i \gamma_j \gamma_5$, where $j=1,2, \cdots, d$. After performing the trace in Eqs.~(\ref{self-energy}) and (\ref{vertex}), we obtain the following contributions from the self-energy and vertex correction in the fermion bubble
\begin{eqnarray}\label{self-energy-current}
\Sigma^{\hat{N}=i{\gamma_j}/i \gamma_j \gamma_5}_{lm}(i\Omega)&=& -\delta_{lm} \frac{8d N_f \Delta_N}{v^{2d-2}} \: \left(\frac{e^2}{h}\right) \int \frac{d^d \mathbf{k}\; d^d \mathbf{p} \; d\omega}{(2 \pi)^{2d}} \:       \frac{\omega(\omega+\Omega)(\omega^2-k^2) - 2(\frac{2}{d}-1) \omega^2 k^2}{((\omega+\Omega)^2+k^2)(\omega^2+k^2)^2 \; (\omega^2+p^2)},  \\
V^{\hat{N}=i{\gamma_j}/i \gamma_j \gamma_5}_{lm}(i\Omega)&=& -\delta_{lm}\frac{4(2-d) N_f \Delta_N}{v^{2d-2}} \: \left(\frac{e^2}{h}\right) \int \frac{d^d \mathbf{k}\; d^d \mathbf{p} \; d\omega}{(2 \pi)^{2d}} \:
 \frac{\left[\omega(\omega+\Omega)-\left( \frac{2}{d}-1 \right) k^2 \right] \left[\omega(\omega+\Omega)-\left( \frac{2}{d}-1 \right) p^2 \right]}{(\omega^2+k^2)\; (\omega^2+p^2) \; \left[ (\omega+\Omega)^2 +k^2 \right] \; \left[ (\omega+\Omega)^2 +p^2 \right]}, \nonumber \\ \label{vertex-current}
\end{eqnarray}
for $N=C, M$ (see Table~\ref{table-disorder}). After completing the integral over frequency and momentum the correction to the polarization bubble is given by
$$\delta\Pi^{\hat{N}=i{\gamma_j}/i \gamma_j \gamma_5}_{lm}(i \Omega)= -\delta_{lm} \: \frac{e^2 N_f\Delta_N}{h} \times \frac{(i\Omega)^3}{24\pi v^4}.$$
\\

Finally we delve into spin-orbit and axial magnetic disorder for which $\hat{N}=\gamma_0 \gamma_j$ and $\hat{N}=i \gamma_0 \gamma_j \gamma_5$, respectively, as shown in Table~\ref{table-disorder}. For these two choices of disorder vertex the contribution from the self-energy and vertex diagrams read as
\begin{eqnarray}
\Sigma^{\hat{N}=\gamma_0\gamma_j/i\gamma_0\gamma_j \gamma_5}_{lm}(i\Omega)&=& -\delta_{lm} \frac{8d N_f \Delta_N}{v^{2d-2}} \: \left(\frac{e^2}{h}\right) \int \frac{d^d \mathbf{k}\; d^d \mathbf{p} \; d\omega}{(2 \pi)^{2d}} \:       \frac{\omega(\omega+\Omega)(\omega^2-k^2) - 2(\frac{2}{d}-1) \omega^2 k^2}{((\omega+\Omega)^2+k^2)(\omega^2+k^2)^2 \; (\omega^2+p^2)},  \\
V^{\hat{N}=\gamma_0\gamma_j/i\gamma_0\gamma_j \gamma_5}_{lm}(i\Omega)&=& \delta_{lm}\frac{4(2-d) N_f \Delta_N}{v^{2d-2}} \: \left(\frac{e^2}{h}\right) \int \frac{d^d \mathbf{k}\; d^d \mathbf{p} \; d\omega}{(2 \pi)^{2d}} \:
 \frac{\left[\omega(\omega+\Omega)-\left( \frac{2}{d}-1 \right) k^2 \right] \left[\omega(\omega+\Omega)-\left( \frac{2}{d}-1 \right) p^2 \right]}{(\omega^2+k^2)\; (\omega^2+p^2) \; \left[ (\omega+\Omega)^2 +k^2 \right] \; \left[ (\omega+\Omega)^2 +p^2 \right]}, \nonumber \\
\end{eqnarray}
for $N=SO, AM$. Correction to the polarization bubble due to these two types of elastic scatterers is given by
$$\delta\Pi^{\hat{N}=\gamma_0\gamma_j/i\gamma_0\gamma_j \gamma_5}_{lm}(i \Omega)= \delta_{lm} \: \frac{e^2 N_f\Delta_N}{h} \times \frac{(i\Omega)^3}{24\pi v^4}.$$

From the correction to the polarization bubble due to disorder we obtain the total imaginary part of the OC to be
\begin{equation}
\Im(\sigma)(\Omega)= \frac{e^2}{h} \; \frac{N_f \Omega}{6 \pi v} \; \left\{ - \log \left[ \frac{4 v^2 \Lambda^2}{\Omega^2} -1\right] + f\left( \Delta_N \right) \; \frac{\Omega \Delta_N}{4 v^3} \right\},
\end{equation}
where $f(x)$ is a \emph{sign} function that takes values $\pm 1$ depending on the nature of the impurity scatterer, as shown in Table~\ref{table-disorder}. The imaginary part of the OC is related to the dielectric function, which now reads as
\begin{equation}
\varepsilon(\Omega)=1-\frac{4 \pi}{\Omega} \Im(\sigma)(\Omega)=1+\frac{2e^2 N_f}{3hv} \; \left\{ \log \left[ \frac{4 v^2 \Lambda^2}{\Omega^2} -1\right] - f\left( \Delta_N \right) \; \frac{\Omega \Delta_N}{4 v^3} \right\}.
\end{equation}

By employing the second Kramers-Kronig relation we can immediately find the correction to the real part of the OC to be
\begin{eqnarray}
\delta \Re(\sigma)(\Omega)= \frac{2}{\pi} \; \mathcal{P} \; \int^{2 v \Lambda}_{0} d \Omega^\prime \; \frac{\Omega^\prime \; \delta \Im(\sigma)(\Omega^\prime)}{{\Omega^\prime}^2-\Omega^2}, \quad \mbox{where} \:\:\: \delta \Im(\sigma)(\Omega)= f(\Delta_N)\;\frac{\sigma_0}{\pi} \times \frac{\Delta_N \Omega}{4 v^3},
\end{eqnarray}
leading to
\begin{eqnarray}
\delta \Re(\sigma)(\Omega)= \sigma_0 \; \left( \frac{\Delta_N \Lambda }{\pi^2 v^2} \right) \; \left[1- \left(\frac{\Omega}{2 v \Lambda} \right) \coth^{-1} \left(\frac{2 v \Lambda}{\Omega}\right) \right] \; f\left(\Delta_N \right).
\end{eqnarray}
Hence the total OC in wekly disordered WSM is given by
\begin{equation}
\Re(\sigma)(\Omega)= \sigma_0 \left[ 1+ \left(\frac{\Delta_N \Lambda}{\pi^2 v^2}\right) \; f\left( \Delta_N \right) + f\left( \Delta_N \right) {\mathcal O} \left( \left[\frac{\Omega}{2 v \Lambda} \right]^2 \right) \right].
\end{equation}
Notice that in three spatial dimensions $\Delta_N \Lambda/(\pi^2 v^2)=\hat{\Delta}_N$ is the dimensionless disorder coupling. Hence, to the leading order in disorder coupling and for small frequency, i.e. $\Omega/(2v\Lambda) \ll 1$, we can compactly write the total OC as
\begin{equation}
\Re(\sigma)(\Omega)= \sigma_0 \left[ 1+ f\left( \Delta_N \right) \: \hat{\Delta}_N \right].
\end{equation}

$\bullet$ \emph{\bf Method of integration.} Finally, we comment on some essential mathematical steps that were used to compute the optical conductivity of clean and weakly disordered WSMs. Notice that integral over internal frequency ($\omega$) can readily be performed using \emph{residue} formula~\cite{arfken}. On the other hand, we here compute the integrals over the internal momentum in spatial dimensions $d=3-\epsilon$, and at the end send $\epsilon \to 0$, closely following the spirit of \emph{dimensional regularization} that manifestly preserves the \emph{gauge invariance}~\cite{veltman, peskin, juricic-vafek-herbut}.
\\

In particular, when we compute the correction to OC due to disorder [from two-loop diagrams (ii) and (iii) in Fig.~\ref{conductivity}] for which the expressions are given in Eqs.~(\ref{self-energy}) and (\ref{vertex}), the strategy is to first perform the integrals over the Matsubara frequency $\omega$ and the momentum ${\bf k}$, and only at the end integrate over the momentum ${\bf p}$. In the last step, it is useful to separately integrate out the part of the integrand that is {\it even} under $p \rightarrow - p$ in $d=3$. Using the fact that this piece is of the form
\begin{equation}
I=\csc(d\pi)\int_0^\infty\, dp\, F(d,p),
\end{equation}
where $d=3-\epsilon$, $\csc (x)=1/\sin(x)$, and the function $F(d,p)$ has the property $F(d,-p)=(-1)^{2d}F(d,p)$. To compute the last integral, let us consider the following integral, which in our case can readily be computed by using the Cauchy theorem~\cite{arfken}
\begin{equation}
\int_{-\infty}^\infty\, dp\, \csc(d\pi)\, F(d,p)= C+\mathcal{O}(\epsilon).
\end{equation}
After splitting this integral in two parts from $-\infty$ to $0$, and  $0$ to $\infty$, substituting $p\rightarrow - p$ in the first term and using the \emph{parity property} of the function $F(d,p)$, we have
\begin{equation}
\int_{-\infty}^\infty dp \csc(d\pi) F(d,p)=\left(1+(-1)^{2d}\right)I=[2+C_1\epsilon+\mathcal{O}(\epsilon^2)]I,
\end{equation}
with $C_1$ as a constant.
We know that the left hand side is of the form $C+ \mathcal{O}(\epsilon)$, and therefore
\begin{equation}
I=C/2 + \mathcal{O}(\epsilon).
\end{equation}
The part that is odd under $ p \to - p$ is easily computed by using some elementary integrations.
\\

$\bullet$ \emph{\bf Derivation of scaling function for conductivity from renormalization group flow equations.} Let us now present the derivation of the scaling function of optical conductivity in the presence of potential/axial disorder couplings from the renormalization group (RG) flow equations of various coupling constants. After evaluating the relevant one-loop diagrams, shown in Fig.~\ref{selfenergy-vertex}, we arrive at the flow equations
\begin{equation}\label{flow-disorder}
\frac{dv}{dl}=v [z-1-\Delta_N], \:\:\: \frac{d\Delta_N}{dl}=-(d-2) \Delta_N + 2 \Delta^2_N,
\end{equation}
for $N=V/A$. The dimensionless disorder coupling goes as $\Delta_N \Lambda/(2 \pi^2 v^2) \to \Delta_N$. Since the flow equations for these two coupling constants are identical due to the underlying chiral symmetry in WSM, from here onward we only focus only on one of these two disorder couplings and set $\Delta_N=\Delta$ for $N=V/A$. The Wilsonian  momentum-shell RG scheme has previously been highlighted in this Supplementary Information. From the second equation it is clear that WSM-metal quantum phase transition takes place at a critical strength of disorder coupling (dimensionless) $\Delta_c=\frac{d-2}{2}=1/2$ for $d=3$. The Fermi velocity can be kept marginal during the procedure of coarse graining if we allow a scale dependent dynamic scaling exponent $z(l)=1+\Delta_N(l)$. Therefore, at the WSM-metal quantum critical point $z=z_c=3/2$ in $d=3$ and the correlation length exponent $\nu=(d-2)^{-1}$. Since under the coarse graining the high energy cut-off $E_\Lambda=v \Lambda$ goes as $E_\Lambda \to e^{-zl} E_\Lambda$, we can arrive at the flow equation of dimensionless energy, defined as $\varepsilon=\Omega/E_\Lambda$
\begin{equation}\label{flow-energy}
\frac{d\varepsilon}{dl}=z(l) \; \varepsilon \:\: \Rightarrow \:\: \frac{d\varepsilon}{dl}= \left[ 1+\Delta(l) \right] \; \varepsilon,
\end{equation}
where $\Omega$ is the frequency and for convenience we set $\hbar=1$.
\\

Since the gauge invariance mandates that the scaling dimension of conductivity is $(2-d)$, the conductivity scales as $\sigma \sim L^{2-d}$, as stated in the main part of the paper. Therefore, to extract the scaling behavior of the OC in different regimes of the phase diagram (such as WSM, quantum critical regime and metal), shown in Fig.~1 of the paper, we need to identify the appropriate RG scale that sets the \emph{infrared} cutoff for the flow of coupling constants. The flow equations for disorder coupling $\Delta$ from Eq.~(\ref{flow-disorder}) and dimensionless energy $\varepsilon$ from Eq.~(\ref{flow-energy}) can readily be solved to obtain~\cite{goswmai-RGsolution}
\begin{equation}\label{RG-solutions}
\Delta(l)=\frac{\Delta_c}{1+ \left[ \frac{\Delta_c}{\Delta_0} -1\right] e^{(d-2)l}}, \quad
\varepsilon(l)=\varepsilon_0 e^{z_c l} \: \left( \frac{\Delta_c}{\Delta_0}\right)^{(z_c-1)\nu} \: \frac{1}{ \left[1+ \left[ \frac{\Delta_c}{\Delta_0} -1\right] e^{(d-2)l}\right]^{(z_c-1)\nu}},
\end{equation}
where the quantities with subscript ``$0$" represent their bare values. In any regime of the phase diagram, there is a competition between three length scales: spatial correlation length ($\xi_l$), temporal correlation length ($\xi_t$) and system size ($L$). As the critical point is approached the former two length scales respectively diverge as $\xi_l \sim |\delta|^{-\nu}$ and $\xi_t \sim |\delta|^{-z \nu}$, where $\delta=1-\Delta_c/\Delta_0$ measures the reduced distance from the critical point.
\\

When $\Delta_0 > \Delta_c$ the disorder coupling in the zero energy limit ($\Omega \to 0$) diverges at the RG scale $\xi_l$, indicating the onset of a metallic phase at stronger disorder ($\Delta_0>\Delta_c$). Knowing the scaling dimension of OC to be $2-d$, we can immediately arrive at the scaling of OC as $\Omega \to 0$ inside the metallic phase to be
\begin{equation}
\sigma_M (\Omega \to 0) \sim \left( \xi_l \right)^{2-d} \: \Rightarrow \: \sigma_M (\Omega \to 0) \sim |\delta|^{\nu (d-2)},
\end{equation}
in accordance with the quoted scaling form in the main part of the paper.
\\

At the critical point $\xi_l$ and $\xi_t$ both diverge, and the infrared cut-off  $e^l$ for the RG flow is set by the energy at which $\varepsilon(l)\sim 1$. Hence, from the solution of $\varepsilon(l)$ we obtain $e^{l} \sim |\Omega|^{-1/z_c}$, when $\Delta_0=\Delta_c$ (notice in our notation $z=z_c$ when $\Delta_0=\Delta_c$). Therefore, the scaling of the OC at the critical point is given by
\begin{equation}
\sigma_Q(\Omega) \sim \left( e^{l} \right)^{2-d} \:\: \Rightarrow \:\: \sigma_Q(\Omega) \sim \Omega^{(d-2)/z_c},
\end{equation}
as $\Omega \to 0$ in agreement with the result announced in the main part of the paper. Such scaling behavior, however, persists over the entire quantum critical regime (see Fig.~1 of the paper).
\\

In the semimetallic side of the the phase diagram $\Delta(l) \to 0$ as $l \to \infty$. Thus, in this regime we can neglect the factor of \emph{unity} in the denominator of the second equation of Eq.~(\ref{RG-solutions}). The relevant RG scale in this regime is again found from the condition $\varepsilon(l) \sim 1$, yielding $e^l \sim \Omega^{-1} |\delta|^{(z_c-1)\nu}$. Therefore, scaling of the OC in the WSM phase for weak disorder is given by
\begin{equation}
\sigma_W(\Omega) \sim \left( e^l \right)^{2-d} \:\: \Rightarrow \:\: \sigma_W (\Omega) \sim \Omega^{d-2} \; |\delta|^{(1-z_c)(d-2)\nu},
\end{equation}
which is also in agreement with the predicted scaling form of OC inside the semimetallic regime of the phase diagram of a dirty WSM. Therefore, from the solution of leading order RG flow equations and proper identification of the infrared curt-off for the RG flow, we can arrive at the leading order scaling behavior of the OC inside the metallic, quantum critical and Weyl semimetal phases.
\\

Within this framework, one can immediately arrive at the scaling behavior (at least the leading oder) of dc conductivity (in the collision dominated regime), simply by redefining the dimensionless energy scales to be $\varepsilon=T/E_0$ (after setting $k_B=1$). Thus, we strongly believe that scaling behavior for both ac and dc conductivity are qualitatively the same in dirty WSMs.
\\

$\bullet$ \emph{\bf Scaling of finite metallic conductivity (as frequency $\Omega \to 0$) with system size.} As shown in the main part of the paper the scaling behavior of optical conductivity is captured by the following universal scaling ansatz
\begin{align}\label{Eq:OC_supple}
\sigma(\Omega,\delta,L) = L^{2-d} {\mathcal G} \left( \frac{L}{\delta^{-\nu}}, \frac{\Omega}{\delta^{\nu z}}\right)= \delta^{\nu (d-2)} {\mathcal F} \left( \frac{L}{\delta^{-\nu}}, \frac{\Omega}{\delta^{\nu z}} \right),
\end{align}
 where ${\mathcal G}$ and ${\mathcal F}$ are two unknown, but universal scaling functions. We now focus on the scaling of the finite metallic conductivity as $\Omega \to 0$ inside the compressible diffusive metallic phase when $\delta>0$ or $\Delta_N>\Delta^{\ast}_N$. Setting $\Omega=0$ in the above equation we obtain
\begin{equation}
\sigma(0,\delta,L)=\delta^{\nu (d-2)} {\mathcal F} \left( L\delta^{\nu}, 0\right).
\end{equation}
Thus upon extracting the finite metallic conductivity for various system sizes and strengths of strong disorder couplings (i.e. when $\delta>0$ or $\Delta_N>\Delta^{\ast}_N$) one can independently extract the correlation length exponent ($\nu$) across the WSM-metal quantum phase transition from the data collapse obtained by comparing $\sigma(0,\delta,L) \delta^{\nu (2-d)}$ vs. $L\delta^{\nu}$ or $L^{1/\nu}\delta$.

\end{document}